\documentclass{elsart}

 \usepackage{epsfig}

\begin{document}

\begin{frontmatter}



\title{Evidence for Solar Neutrino Flux Variability and its Implications}

\author[doc]{D.O.~Caldwell}
\author[pas]{and P.A.~Sturrock}
\address[doc]{Physics Department, University of California\\ Santa Barbara,
CA 93106-9530, USA}
\address[pas]{Center for Space Science
        and Astrophysics, Stanford University, Stanford, CA 94305-4060, USA}

\begin{abstract}
Although KamLAND apparently rules out
Resonant-Spin-Flavor-Precession (RSFP) as an explanation of the
solar neutrino deficit, the solar neutrino fluxes in the Cl and Ga
experiments appear to vary with solar rotation.  Added to this
evidence, summarized here, a power spectrum analysis of the
Super-Kamiokande data reveals significant variation in the flux
matching a dominant rotation rate observed in the solar magnetic
field in the same time period.  Three frequency peaks, all related
to this rotation rate, can be explained quantitatively.  A
Super-Kamiokande paper reported no time variation of the flux, but
showed the same peaks, there interpreted as statistically
insignificant, due to an inappropriate analysis.  This modulation is
small (7\%) in the Super-Kamiokande energy region (and below the
sensitivity of the Super-Kamiokande analysis) and is consistent with
RSFP as a subdominant neutrino process in the convection zone. The
data display effects that correspond to solar-cycle changes in the
magnetic field, typical of the convection zone.  This subdominant
process requires new physics: a large neutrino transition magnetic
moment and a light sterile neutrino, since an effect of this
amplitude occurring in the convection zone cannot be achieved with
the three known neutrinos. It does, however, resolve current
problems in providing fits to all experimental estimates of the mean
neutrino flux, and is compatible with the extensive evidence for
solar neutrino flux variability.
\end{abstract}
\end{frontmatter}

\newpage

\section{Introduction}
Results from the KamLAND experiment \cite{ref:1} seem to confirm the
Large-Mixing Angle (LMA) solution to the solar neutrino deficit and
rule out the Resonant-Spin-Flavor-Precession (RSFP) explanation
\cite{ref:2}. On the other hand, there is increasing evidence
\cite{ref:3,ref:4,ref:5,ref:6,ref:7,ref:8,ref:9,ref:9a} that the
solar neutrino flux is not constant as assumed for the LMA solution,
but varies with periods that can be attributed to well-known solar
processes. This suggests that the solar neutrino situation may be
more complex than is usually assumed, and that RSFP may be
subdominant to LMA \cite{ref:10,ref:16new}, requiring a large
transition magnetic moment and hence new physics. Since this recent
information on solar neutrino variability is not widely known, a
brief summary is presented here of analyses of radiochemical
neutrino data, along with newer input from the Super-Kamiokande
experiment \cite{ref:11}. Although the 10-day averages of
Super-Kamiokande solar neutrino data \cite{ref:12} show no obvious
time dependence, power-spectrum analyses
\cite{ref:13a,ref:14b,ref:14c} displayed a strong peak at the
frequency $26.57\pm0.05$ y$^{-1}$ (period 13.75 d), as well as one
at 9.42 y$^{-1}$.  These were clearly an alias pair, due to the
extremely regular 10-day binning for which the timing had a strong
periodicity with frequency 35.99 y$^{-1}$ ($=26.57+9.42$).  A
subsequent Super-Kamiokande paper \cite{ref:14a} provided 5-day
averages of the data, and there was no longer evidence for the 26.57
y$^{-1}$ peak, showing it to be an alias of the 9.42 y$^{-1}$ peak
(which will be explained later) due to the extremely regular 10-day
binning. In the 5-day data the 9.42 y$^{-1}$ is enhanced.  We also
find a peak at 39.28 y$^{-1}$, which is the frequency of a dominant
rotation-related oscillation in the photospheric magnetic field.
Another notable peak at 43.72 y$^{-1}$ may be attributed to the same
physical process as that responsible for the peak at 9.42 y$^{-1}$.

A suggested subdominant process that is compatible with these
periodicities involves an RSFP transition in which the solar $\nu_e$
changes to a different flavor, and the spin is flipped.  As in the
MSW process \cite{ref:23}, this can occur resonantly at an
appropriate solar density.  The two processes, LMA MSW and RSFP,
would take place sequentially at different solar radii. If the RSFP
process were to be achieved with the three known active neutrinos,
their measured mass differences would require that RSFP occurs at a
smaller solar radius than that at which the MSW effect takes place,
placing it in the solar core.  This case has been analyzed
\cite{ref:16new}, with the result that the flux modulation produced
by RSFP is very small, varying with neutrino energy from 0.8\% at
2.5 MeV to 4\% at 13 MeV for a product of field and magnetic moment
of $10^{-6}$G$\mu_B$.


Another model was suggested in an earlier version of this paper, and
predictions from it have been calculated \cite{ref:10}.  This model
utilizes a sterile neutrino that couples to the electron neutrino
only through a transition magnetic moment.  The lack of mixing with
active neutrinos avoids all known limitations on sterile neutrinos,
and the sterile final-state also makes irrelevant the usual
constraints on RSFP from the null observations of solar
antineutrinos.  The solar data require a mass-squared difference
between the electron and sterile neutrinos of $\Delta
m^2\sim10^{-8}$ eV$^2$.  This is different from the mechanism and
the sterile state suggested by de~Holanda and Smirnov \cite{ref:16a}
for a subdominant transition to improve agreement with the Homestake
data \cite{ref:15} and the Super-Kamiokande energy dependence
\cite{ref:11}, but again a decided improvement in the fit to the
solar neutrino data is obtained \cite{ref:10}.  This improvement,
and also the reason RSFP by itself actually provides a better fit to
mean solar data than does LMA \cite{ref:13,ref:14}, results from the
shape of the RSFP neutrino survival probability.  It has a resonance
pit at a density that suppresses the 0.86 MeV $^7$Be line (as does
the Small-Mixing Angle [SMA] solution), but tends toward 1/2 and
hence fits the Super-Kamiokande spectrum, whereas the survival
probability goes to unity in the SMA case. The high-energy rise
includes the neutral current scattering of the products of the
spin-flavor flips, which need to be active, and hence Majorana
neutrinos are required in order to fit \cite{ref:14} the SNO data
\cite{ref:18a} for RSFP by itself.  In the case of subdominant RSFP,
the high-energy behavior is determined mainly by the MSW transition,
since the observed flux modulation is found to be small ($\sim7$\%)
in the $^8$B neutrino region.  Going down to intermediate energies,
the dip toward the resonance pit reduces the predicted rate for the
Homestake experiment \cite{ref:15}, improving agreement with the
data, and eliminates the rise below $\sim8$ MeV which is predicted
by the LMA solution but not observed by Super-Kamiokande
\cite{ref:11} or SNO \cite{ref:18a}.  The
electron-neutrino-associated sterile neutrino, which provides a
better fit to the solar data and is required for this version of
subdominant RSFP, is different from the muon-neutrino-associated
sterile neutrino needed for the results of the LSND experiment
\cite{ref:15a}.  Possibly three sterile neutrinos exist with a
family symmetry.

\section{Review of Past Evidence}
The three-neutrino and four-neutrino RSFP models discussed above
differ in their predictions of the location of the RSFP process in
the Sun and in the size of the flux modulation they produce.  The
modulation is larger in the case with the added sterile neutrino,
and the modulation can increase at smaller energies \cite{ref:10},
the reverse of the other model.  These distinctions should be kept
in mind as the previously published evidence for solar neutrino flux
variability is reviewed briefly.

By analyzing $10^3$ simulated data sequences, it was found
\cite{ref:3} that the variance of the Homestake solar neutrino data
\cite{ref:15} is larger than expected at the 99.9\% confidence level
(CL).  A power spectrum analysis \cite{ref:3} of the data showed a
peak at $12.88\pm0.02$ y$^{-1}$ (28.4 d), compatible with the
rotation rate of the solar radiative zone.  Peak widths are computed
for a probability drop to 10\%.  A lower limit to the width of a
peak is set by the duration of the time series---in this case 24 y,
so that the power spectrum analysis should be able to resolve peaks
with separation as small as 0.04 y$^{-1}$.  Four sidebands to the
12.88 y$^{-1}$ frequency gave evidence at the 99.8\% CL for a
latitudinal effect associated with the tilt of the sun's rotation
axis, and the latitude dependence was also seen directly in the data
at the 98\% CL \cite{ref:4}.

\begin{figure}[b]
\epsfig{file=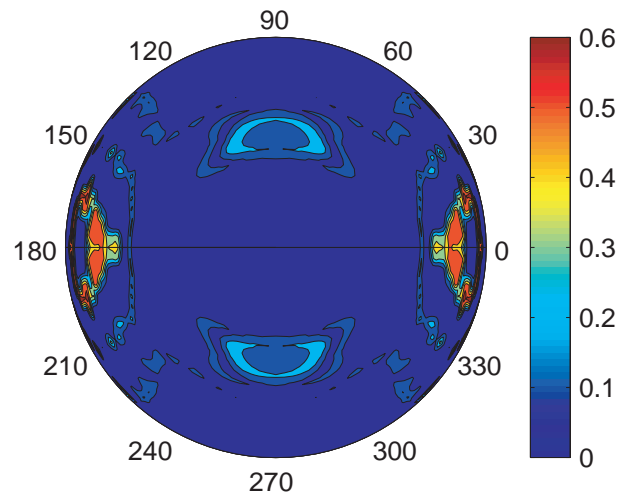} \caption{Map of the resonance statistic
of the SOHO/MDI helioseismology and GALLEX data on a meridional
section of the solar interior.  The only high probability areas
(red) are lens-shaped sections near the equator, and all others are
low probability. \label{fig:1}}
\end{figure}

The GALLEX data \cite{ref:16} showed a peak at $13.59\pm 0.04$
y$^{-1}$, compatible with the equatorial rotation rate of the deep
convection zone \cite{ref:7,ref:8}.  This peak is also in the
Homestake data, and a combined analysis of both datasets shows that
the 13.59 y$^{-1}$ peak is stronger than in either dataset alone.
Comparison \cite{ref:7} of the power spectrum for the GALLEX data
with a probability distribution function for the solar synodic
rotation frequency as a function of radius and latitude, derived
from SOHO/MDI helioseismology data \cite{ref:17}, results in
Fig.~\ref{fig:1}.  This map shows that the rotation frequency
matches the neutrino modulation in the equatorial section of the
convection zone at about 0.8 of the solar radius, $R_\odot$.  Note
that this is compatible with the RSFP model that involves a sterile
neutrino, but not with the three-neutrino model.

\begin{figure}
\epsfig{file=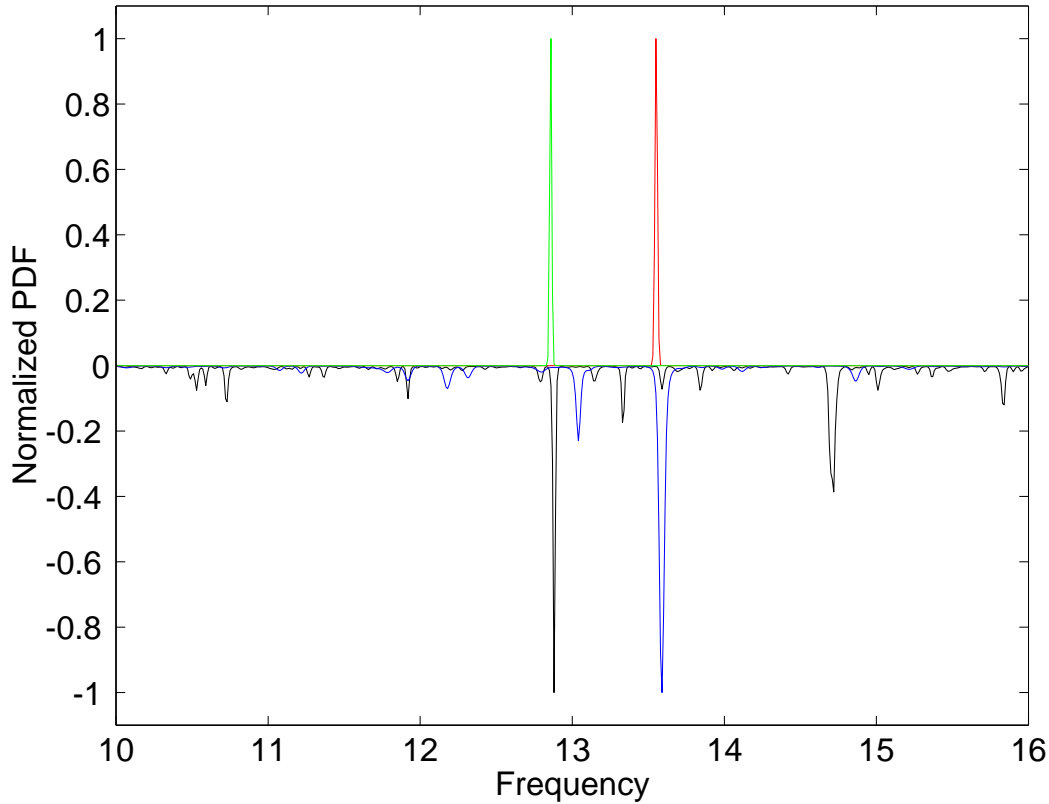,width=5.5in} \caption{Comparison of
normalized probability distribution functions formed from power
spectra of data from SXT equator (red), SXT N60-S60 (green),
Homestake (black), and GALLEX (blue).  Note that the SXT (red) and
GALLEX data are equatorial, and the other two are not.  In this and
all subsequent figures frequencies are in cycles per year.
\label{fig:2}}
\end{figure}

The influence of these rotation frequencies extends even to the corona,
since the SXT instrument \cite{ref:18} on the Yohkoh spacecraft provides
X-ray evidence for two ``rigid'' rotation rates, one ($13.55\pm0.02$ y$^{-1}$)
mainly at the equator, and the other ($12.86\pm0.02$ y$^{-1}$) mainly at high
latitudes.  These values are in remarkable agreement \cite{ref:8} with the neutrino modulation
frequencies and also with their equatorial location in one case and non-equatorial
in the other; see Fig.~\ref{fig:2}.
A plausible interpretation of this result is that a large-scale magnetic structure
deep in the convection zone modulates the neutrino flux and is also responsible for
the structure of the low-$k$ component of the photospheric magnetic field (where $k$
is wave-number), and that the coronal magnetic field reflects the low-$k$ components
rather than the high-$k$ components that arise from granulation.

The fact that coronal X-rays and the neutrino flux both show
evidence of two dominant rotation frequencies is quite consistent
with similar results of analyses of other solar variables.  For
instance, an analysis \cite{ref:19} of the photospheric magnetic
field during solar cycle 21 found two dominant magnetic regions: one
in the northern hemisphere with synodic rotation frequency
$\sim13.6$ y$^{-1}$, and the other in the southern hemisphere with
synodic rotation frequency $\sim13.0$ y$^{-1}$.  Similarly, an
analysis \cite{ref:20} of flares during solar cycle 23 found a
dominant synodic frequency of $\sim13.5$ y$^{-1}$ for the northern
hemisphere and $\sim12.9$ y$^{-1}$ for the southern hemisphere.
These and other studies show a strong tendency for magnetic
structures to rotate either at about 12.9 y$^{-1}$ or at about 13.6
y$^{-1}$.  Since the Sun's magnetic field is believed to originate
in a dynamo process at or near the tachocline, it is possible that
some magnetic flux is anchored in the radiative zone just below the
tachocline, where the synodic rotation frequency is about 12.9
y$^{-1}$, and some just above the tachocline in the convection zone,
where the synodic rotation frequency is about 13.6 y$^{-1}$.  It is
also possible that the 12.9 y$^{-1}$ frequency results from a
latitudinal wave motion in the convection zone excited by structures
at or near the tachocline.

An example of latitudinal oscillatory motion of magnetic structures
may be the well-known Rieger-type oscillations (seen in gamma-ray
flares, sunspots, etc.) with frequencies of about 2.4, 4.7, and 7.1
y$^{-1}$ \cite{ref:21}.  These periodicities may be attributed to
r-mode oscillations \cite{ref:31} with spherical harmonic indices
$\ell=3$, $m=1$,2,3, giving frequencies $\nu=2m\nu_R/\ell(\ell+1)$,
where $\nu_R$ is the sidereal rotation rate $[\nu({\rm
sid})=\nu({\rm syn})+1]$.  These frequencies are also seen in
neutrino data \cite{ref:3,ref:5}. A joint spectrum analysis
\cite{ref:9a} of Homestake and GALLEX-GNO data yields peaks at
12.88, 2.33, 4.62, and 6.94 y$^{-1}$, indicating a sidereal rotation
frequency $\nu_R = 13.88\pm 0.03$ y$^{-1}$.
These r-mode frequencies provide further, and somewhat unexpected,
evidence for neutrino flux variations, and we return later to another observation of them.

\section{Observations in the Neutrino Context}
The difference between the main modulations detected by the Cl and
Ga experiments may be explained by the tilt of the solar axis
relative to the ecliptic, along with the fact that Cl and Ga
neutrinos are produced mainly at quite different radii.  The Ga data
come mostly---especially in the sterile-neutrino model, where the
fit requires suppressing the $^7$Be line---from $pp$ neutrinos,
which originate predominantly at large solar radius ($\sim0.2$
R$_\odot$), so that the wide beam of neutrinos detected on Earth is
insensitive to axis tilt.  Thus the beam of neutrinos detected by
the Ga experiments exhibits no seasonal variation and can be
modulated by the equatorial structure indicated in Fig.~\ref{fig:1},
leading to the observed frequency at about 13.6 y$^{-1}$. On the
other hand, the Homestake experiment detects neutrinos produced from
a smaller sphere ($\sim0.05$ R$_\odot$), so that the axis tilt may
cause these neutrinos mainly to miss the equatorial structure of
Fig.~\ref{fig:1} and instead sample nonzero latitudes where the 12.9
y$^{-1}$ modulation may be more significant.  Twice a year, axis
tilt has no effect for these neutrinos, leading to a seasonal
variation in the measured flux \cite{ref:4}.

In the sterile-neutrino model, the 13.6 y$^{-1}$ frequency, located as in
Fig.~\ref{fig:1}, represents a modulation
of the $pp$ neutrinos which are at or near the steeply falling edge of the neutrino
survival probability as it dips toward the RSFP resonance pit.  This pit, which suppresses
the $^7 \rm{Be}$ neutrinos, is where the largest value of the mass-squared difference
between the initial and final states, divided by their energy, satisfies
\begin{equation}
\Delta m^2/E=2\sqrt 2G_FN_{\rm eff},
\label{eq:1}
\end{equation}
and is essentially the same as for an MSW resonance \cite{ref:23},
since it also arises from a difference in forward-scattering
amplitudes of the two states.  Ignoring angle factors, for MSW
$N_{\rm eff}=N_e$ (the electron density), and for RSFP and
active-active transitions, $N_{\rm eff}=N_e-N_n$ for Majorana
neutrinos, where $N_n$ is the neutron density (about $N_e/6$ in the
region of interest in the Sun).  For Dirac neutrinos $N_{\rm
eff}=N_e-N_n/2$, but for RSFP alone (not the subdominant case), only
Majorana neutrinos fit all the solar data.  For the more complicated
subdominant case involving a transition magnetic moment coupling to
a sterile state, the numerical result of Ref.~\cite{ref:10} may be
used.  It shows the resonant transition occurring close to 0.8
R$_\odot$ for $\Delta m^2=1.8\times10^{-8}$ eV$^2$, which provides a
good fit to the data. This result is in remarkable agreement with
the position shown in Fig.~\ref{fig:1}.

\section{Changes with Solar Cycle}
It is well known that the convection zone magnetic field changes with the solar
cycle, so the neutrino modulation features described above should not be permanent
in the sterile-neutrino model.
It has been argued \cite{ref:24} that an RSFP effect would have to be in the
radiative zone (the location of the RSFP in the three-neutrino model), where the
field does not change with the solar cycle, on the assumption that solar
cycle variations are not manifested.  On the contrary, we show here that solar
cycle changes play an important role.  Variations in neutrino rates are difficult to
observe, since changes in field magnitude would be undetectable if the transition
remains adiabatic, and even if flux modulation results, average rates may vary
only slightly.

\begin{figure}
\epsfig{file=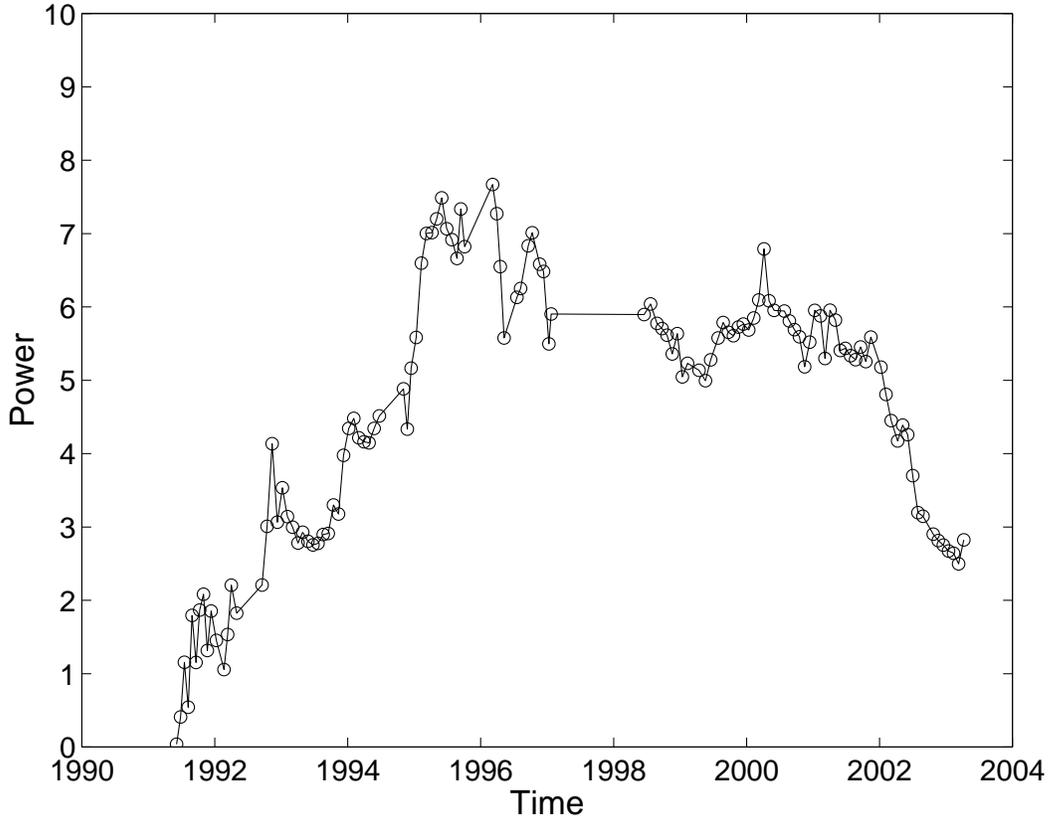,width=5.5in} \caption{Running Rayleigh
power, as a function of end time, for frequency 13.59 y$^{-1}$ from
the GALLEX-GNO data.  Note that the power builds up from the start
of data taking after the 1990 solar maximum until the 1996.8 solar
minimum, after which there is little evidence for that frequency.
\label{fig:3}}
\end{figure}

In the sterile-neutrino model the feature that is most sensitive to
field magnitude or radial variation is the intersection of the very
steeply falling $pp$ neutrino spectrum with the edge of the RSFP
resonance pit.  The shape of the neutrino transition probability
function depends not only on the resonance location,
Eq.~(\ref{eq:1}), but also on the product of the magnetic moment and
magnetic field strength. Thus a change in field will shift the
intersection of the $pp$ spectrum and resonant pit edge and change
the degree of modulation. As can be seen in Fig.~\ref{fig:3}, the
peak at 13.6 y$^{-1}$ associated with these neutrinos increased in
strength from the start of data taking after the solar maximum of
1989.6 to the solar minimum of 1996.8, after which the modulation
becomes weak. Also it was during that cycle that the main buildup in
the strength of the 12.9 y$^{-1}$ oscillation was detected by
Homestake.  The SXT X-ray data, with which these two frequencies
have remarkable agreement \cite{ref:8}, also came from that same
solar cycle.

In response to a paper \cite{ref:31a} suggesting that the 13.59
y$^{-1}$ line is not statistically significant in the GALLEX-GNO
data, we have recently shown \cite{ref:32a} that the difference is
due to different choices made in the analysis procedures, such as
whether or not one takes into account solar cycle changes. Using the
most sensitive analysis choices, we found the 13.59 y$^{-1}$
modulation to be significant at the 99.9\% confidence level.

\begin{figure}
\epsfig{file=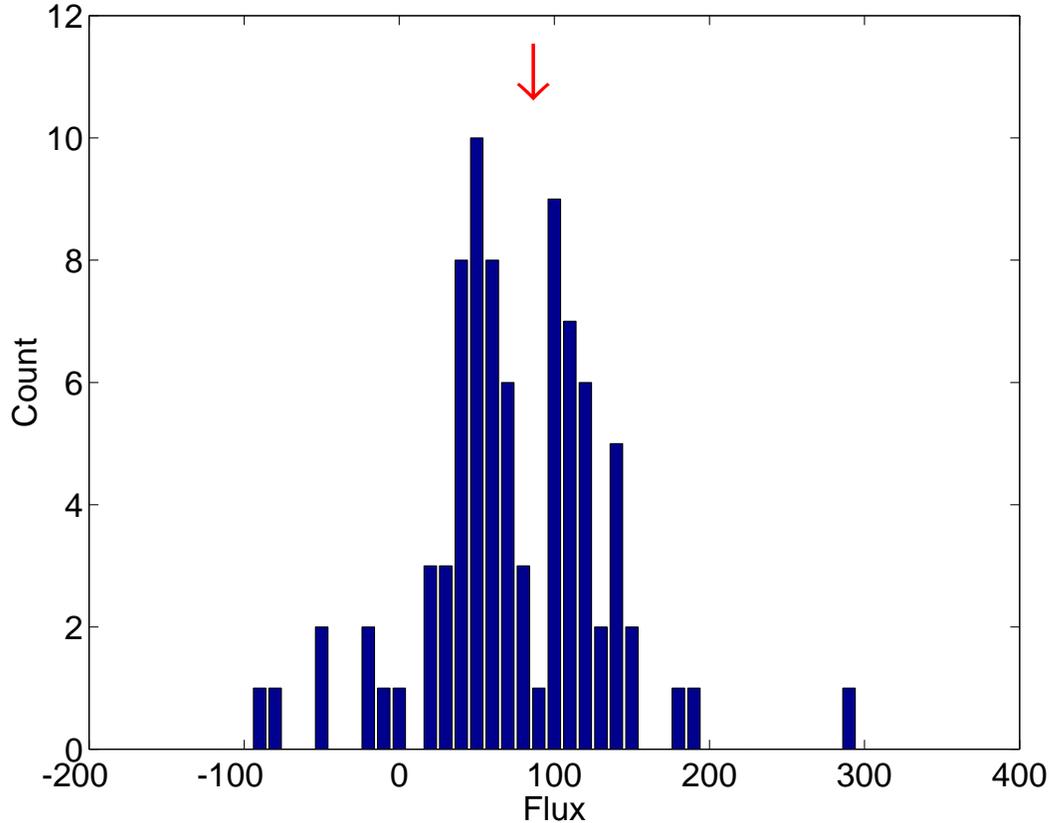,width=5.5in} \caption{Histogram of
GALLEX data.  The arrow indicates the maximum-likelihood estimate of
the flux (in SNU). \label{fig:4}}
\end{figure}

\begin{figure}
\epsfig{file=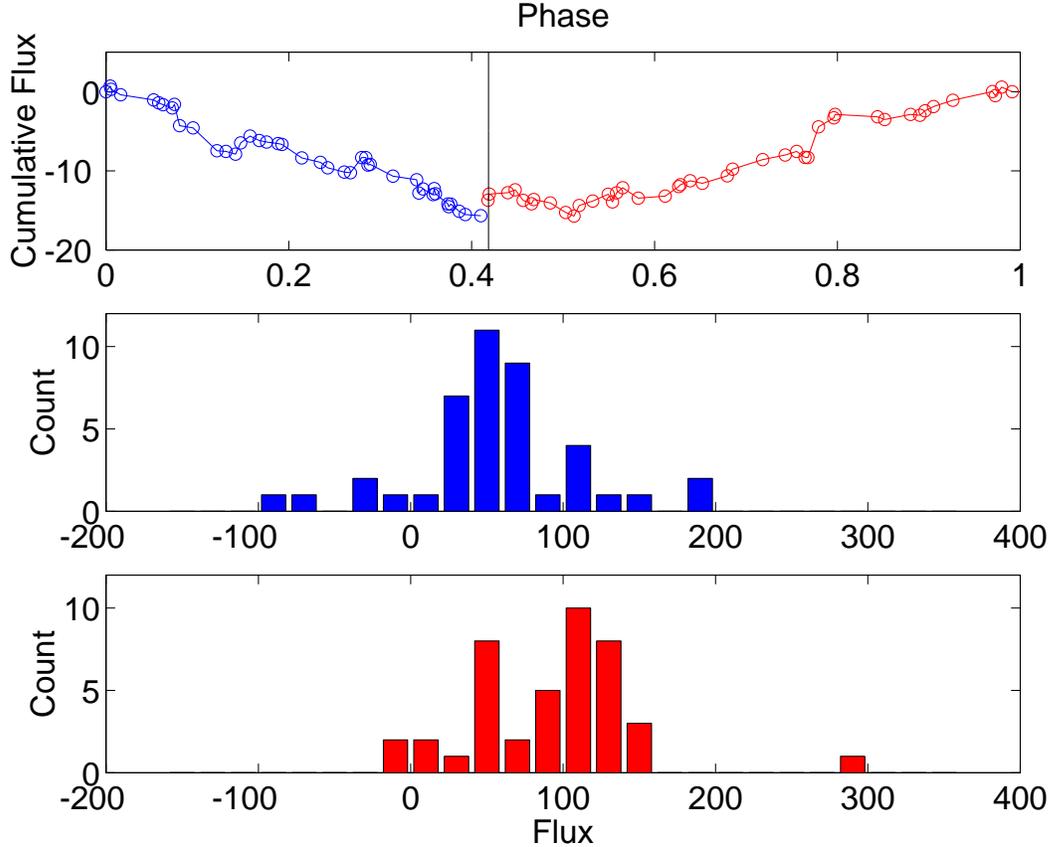,width=5.5in} \caption{Normalized
GALLEX flux (in SNUs) measurement runs (prior to 1997) reordered
according to the phase of the 13.59 y$^{-1}$ solar modulation.  The
division in phase is made so as to have equal numbers of events in
the descending (blue) and ascending (red) parts of the cycle.
\label{fig:5}}
\end{figure}

Another feature one may attribute to a time variation of the
intersection of the $pp$ spectrum with the edge of the RSFP
resonance pit is the appearance \cite{ref:6} of a bimodal flux
distribution in the Ga data, as shown in Fig.~\ref{fig:4}.  A single
peak would be expected if there were no flux modulation.  This
structure is clearly evident in unbinned data during the same solar
maximum to solar minimum period.  This neutrino flux effect also
diminishes after the solar minimum. Adding to the significance of
this result is the plot of Fig.~\ref{fig:5}, which shows that when
the end times of runs are reordered according to the phase of the
13.59 y$^{-1}$ rotation, the flux values are low in one-half of the
cycle and high in the other.

It is unfortunate that the Homestake experiment, which detected
mainly the $^8$B neutrinos (the only neutrino component registered
by Super-Kamiokande) ceased operation before Super-Kamiokande
started. As a result, there is no way to predict from other solar
neutrino experiments what neutrino flux variation should have been
detected by Super-Kamiokande during its operation from May 1996
(near solar minimum) until July 2001 (near solar maximum).  However,
it is possible to compare Super-Kamiokande measurements with other
solar variables that reflect the influence of the Sun's internal
magnetic field.  It is important to emphasize that the change in the
solar field at a change in solar cycle makes it impossible to
predict, even for the same experiment, what modulation frequencies
will be observed in a new solar cycle if one has data only from
previous solar cycles.  This would not be true for the
three-neutrino model for which the RSFP process is located in the
stable radiative zone.

\section{Super-Kamiokande Data Analysis}
The Super-Kamiokande group released \cite{ref:12} flux measurements
in 184 bins of about 10 days each. The flux measurements vary over
the range $2:1$, and the fractional error of the measurement
(averaged over all bins) is $0.14$. Because of the regularity of the
binning, the ``window spectrum" (the power spectrum of the
acquisition times) has a huge peak (power $S>120$) at a frequency
$\nu=35.99$ y$^{-1}$ (period 10.15 d).  (Note that the probability
of obtaining a power of strength $S$ or more by chance at a
specified frequency \cite{ref:25} is $e^{-S}$.) This regularity in
binning inevitably leads to severe aliasing of the power spectrum of
the flux measurements, producing consequences explored below.

Analysis of the data in 10-day bins by a likelihood procedure
\cite{ref:26} outlined in the Appendix, in which the start and stop
times for each bin were used so as to take into account the time
over which data were taken, gave \cite{ref:13a} the strongest peak
in the range 0 to 100 y$^{-1}$ at $\nu=26.57$ y$^{-1}$ with
$S=11.26$, and the next strongest peak at $\nu=9.42$ y$^{-1}$ with
$S=7.29$. Milsztajn \cite{ref:14b}, and subsequently the
Super-Kamiokande group \cite{ref:14c}, used the Lomb-Scargle
analysis method \cite{ref:25}, which assigns data acquisition to
discrete times (in this case chosen to be the mid-time of each bin)
rather than to extended intervals. This ``delta-function" form for
the time window function is inappropriate, since the length of each
bin is comparable with the periods of some of the oscillations of
interest. The conventional form of the Lomb-Scargle method, as used
by Milsztajn \cite{ref:14b} and the Super-Kamiokande group
\cite{ref:14c}, assigns equal weight to all data points, whereas a
later development by Scargle \cite{ref:25} provides for non-uniform
weighting as is appropriate if the uncertainties in measurements
vary from bin to bin, and they certainly do in the Super-Kamiokande
data. The likelihood method is superior because it can in principle
incorporate any uncertainty distributions on a bin-by-bin basis as
well as take account of the start time, end time, and mean live time
of each bin.

The Lomb-Scargle analysis by the Super-Kamiokande group
\cite{ref:14c} gave $S = 10.7$ at $\nu=26.55$ y$^{-1}$, which was
stated to correspond to 98.6\% CL, while the ignored peak at 9.42
y$^{-1}$ had $S = 6.3$. Subsequently, the Super-Kamiokande group
modified their analysis to take account of the non-uniformity of
data collection, assigning the flux measurements for each bin to a
``mean live time" rather than the mid-time. This resulted in a
reduction to $S = 7.51$ of the power at 26.55 y$^{-1}$, but an
increase to 6.67 in the power of the peak at 9.42 y$^{-1}$.

\begin{figure}
\epsfig{file=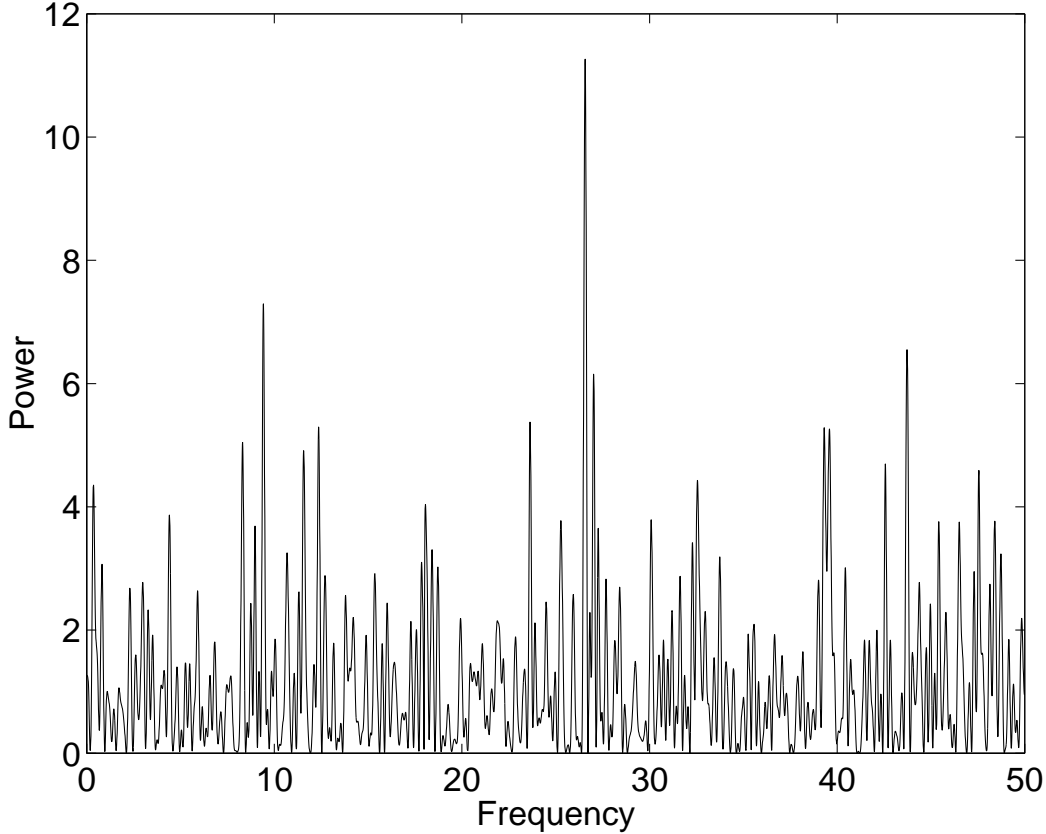,width=5.5in} \caption{Power Spectrum of
Super-Kamiokande 10-day data computed by the likelihood method.
\label{fig:6}}
\end{figure}

We have modified our likelihood analysis to take account of the
non-uniformity in data acquisition, dividing each bin into two parts
(before and after the mean live time) with appropriate weightings.
In contrast to the sensitivity of the Lomb-Scargle procedure to
non-uniformity in the data acquisition process, we find the
likelihood procedure to be quite insensitive to the non-uniformity
represented by the Super-Kamiokande data.  The ``single boxcar" and
``double boxcar" time window functions yield essentially the same
power spectrum, that shown in Figure~\ref{fig:6}, in which the peaks
at 26.55 y$^{-1}$ and at 9.42 y$^{-1}$ still have $S = 11.26$ and
7.29, respectively.  This again shows the superiority of the
likelihood method over the Lomb-Scargle procedure used, since the
latter employed only the mean time, whereas the former utilized
mean, start, and stop times, and more importantly, the measurement
errors.

We previously noted \cite{ref:13a} that 9.42 y$^{-1}$ and 26.57
y$^{-1}$ sum to 35.99 y$^{-1}$, which is the sampling frequency,
from which we inferred that one peak is spurious, being an alias of
the other. Since the peak at 26.57 y$^{-1}$ had the bigger power,
and since it falls in the band of twice the synodic rotation
frequency, we assumed that the peak at 9.42 y$^{-1}$ was merely an
alias. However, when the Super-Kamiokande data were subdivided into
5-day bins \cite{ref:14a}, the peak at 26.57 y$^{-1}$ disappeared,
but the peak at 9.42 y$^{-1}$ remained.  Hence it was clear that the
peak at 26.57 y$^{-1}$ was an alias of the peak at 9.42 y$^{-1}$,
not the other way around.

\begin{figure}
\epsfig{file=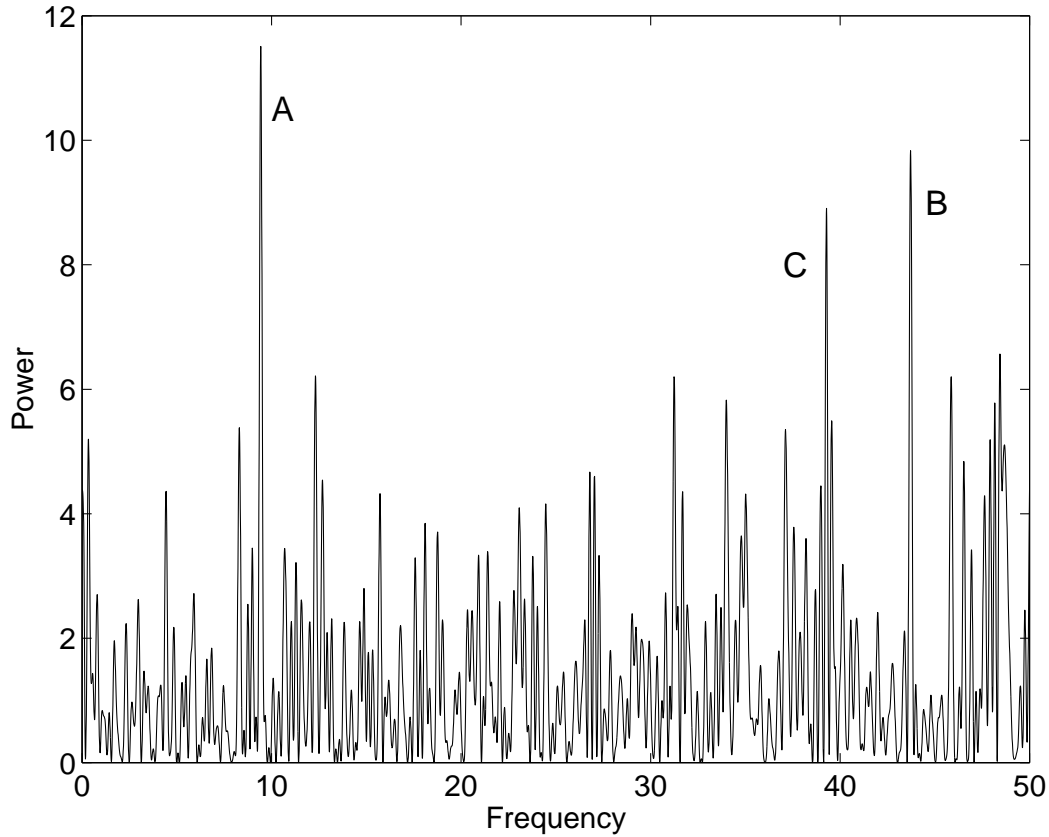,width=5.5in} \caption{Power Spectrum of
Super-Kamiokande 5-day data computed by the likelihood method,
taking account of measurement error, start time, and end time, of
each bin. \label{fig:7}}
\end{figure}

We analyzed the new Super-Kamiokande 5-day data using the likelihood
procedure, taking account of the experimental statistical error
estimates, the mean live time and the start and end time of each bin
\cite{ref:26}. The resulting power spectrum is shown in
Figure~\ref{fig:7}. The biggest peak, A, in the range 0--100
y$^{-1}$ is that at $9.43\pm0.05$ y$^{-1}$, which now has the
enhanced power $S = 11.51$. In agreement with the analysis of the
Super-Kamiokande group, we find that the peak at 26.57 y$^{-1}$ has
disappeared. Hence 9.43 y$^{-1}$ is the primary peak, and the 26.57
y$^{-1}$ peak was merely an alias in the 10-day data.  Since the
5-day binning is also very regular, the power spectrum of the timing
now has a huge peak at 72.01 y$^{-1}$ (period 5.07 days), so an
alias peak would be expected at $72.01-9.43=62.58$ y$^{-1}$, and we
do indeed find a peak (not shown in Fig.~\ref{fig:7}) at
$\nu=62.56\pm0.08$ y$^{-1}$ with $S=5.36$. There are also notable
peaks at $39.28\pm0.05$ y$^{-1}$ ($S=8.91$), labeled C, and
$43.72\pm0.06$ y$^{-1}$ ($S=9.83$), labeled B in Fig.~\ref{fig:7}.
The latter two peaks, plus that at 9.43 y$^{-1}$, are the three
strongest peaks in the range 0--100 y$^{-1}$.  Furthermore, the
three peaks A, B, and C are all stronger in the 5-day dataset than
in the 10-day dataset, as one would expect of real modulation,
because of the difference in the sampling time.

\section{Explanation for the Peak Frequencies}
\begin{figure}
\epsfig{file=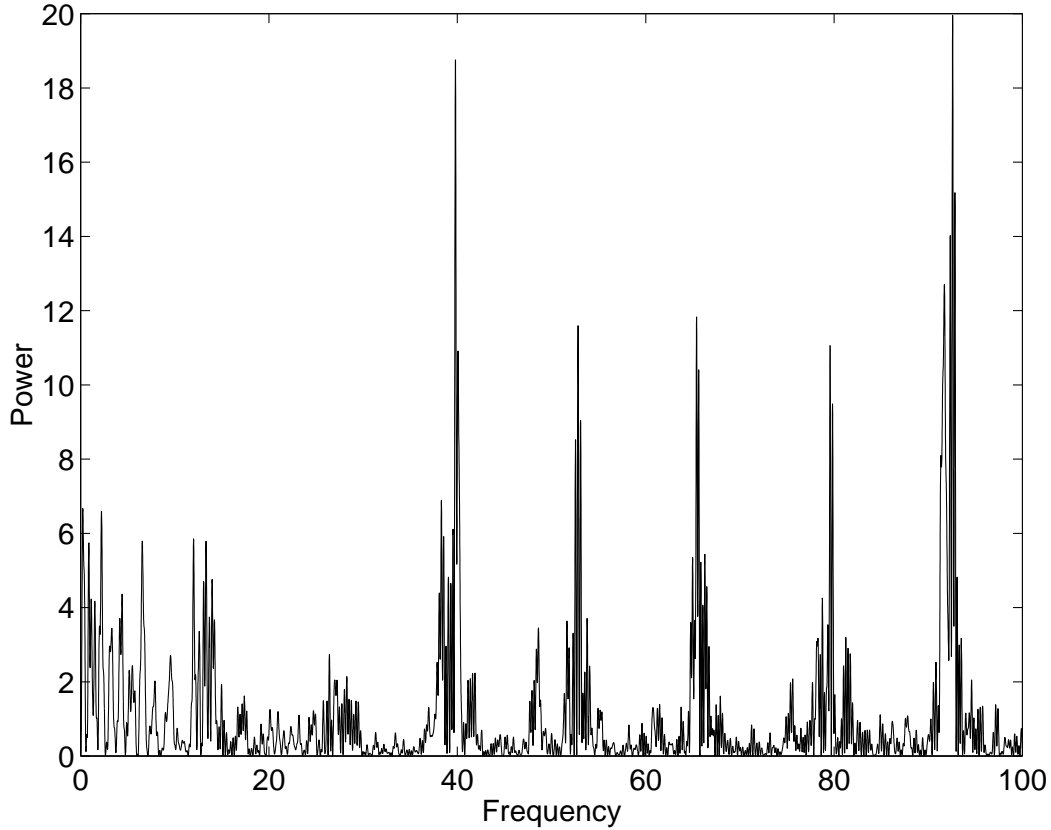,width=5.5in} \caption{Power spectrum of
the disk-center solar magnetic field for the interval of operation
of Super-Kamiokande. \label{fig:8}}
\end{figure}

To understand these three peaks, it is helpful to seek guidance from measurements of the
solar magnetic field.  A power spectrum out to 120 y$^{-1}$ of the photospheric field at
Sun-center during the Super-Kamiokande data-acquisition interval is shown in
Figure~\ref{fig:8}. We see that the fundamental and second harmonic of the rotation
frequency are virtually absent, but there is a remarkable sequence of higher harmonics.
By forming the combined spectrum statistic \cite{ref:9} of the third through the seventh
harmonics, we obtain the estimate $13.20\pm0.14$ y$^{-1}$ for the synodic rotation
frequency of the photospheric magnetic field, leading to the value $14.20\pm0.14$
y$^{-1}$ for the sidereal rotation frequency.

\begin{figure}
\epsfig{file=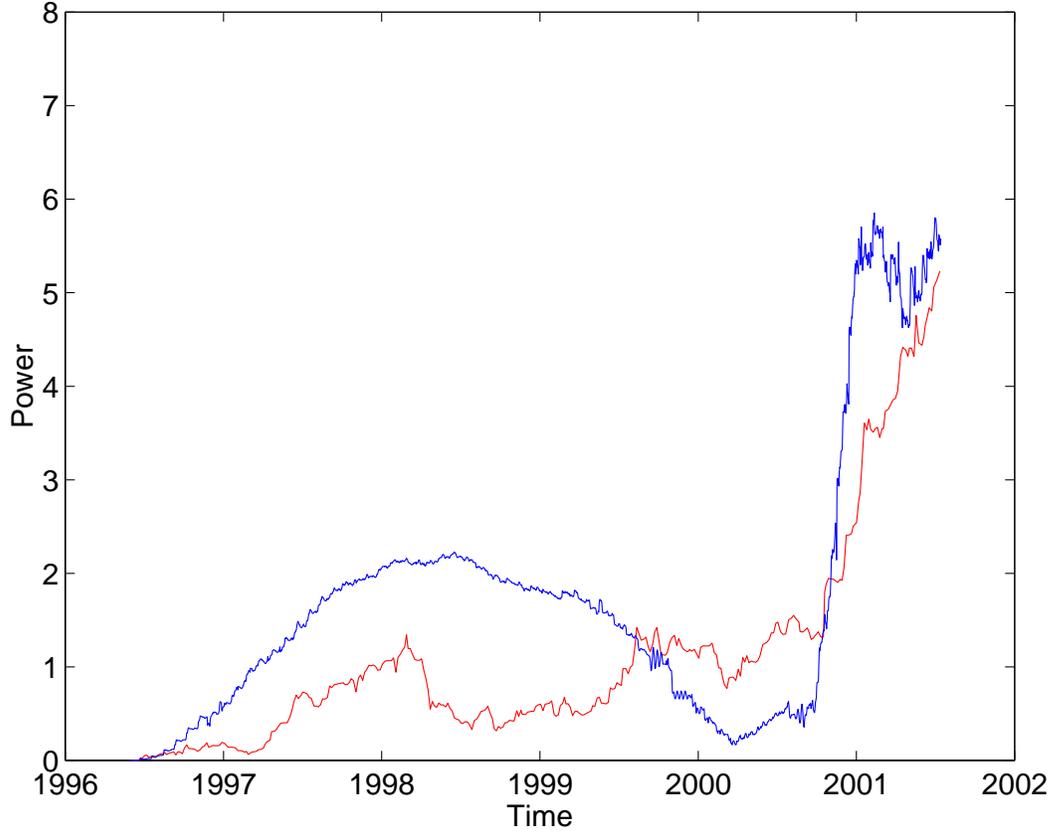,width=5.5in} \caption{Cumulative
contribution to the final Rayleigh power of disk-center magnetic
field (blue) and 5-day Super-Kamiokande data (red) for frequency
39.56 y$^{-1}$. The steep rise begins near solar maximum (2000.6).
\label{fig:9}}
\end{figure}

We see that peak C, at 39.28 y$^{-1}$, falls within the band of
frequencies, $39.60\pm0.42$ y$^{-1}$, of the third harmonic of the
synodic rotation frequency of the magnetic field. We have applied
the shuffle test \cite{ref:3,ref:30}, randomly re-assigning flux and
error measurements (kept together) to time bins.  For this purpose,
the shuffle test is more reliable than Monte Carlo simulations,
since by using actual data there is no need to assume some form for
the probability distribution function of the simulated data. In this
way we find that only 5 of 1,000 simulations yield a power larger
than 8.91 (the power of peak C) in this search band. It appears,
therefore, that peak C may be attributed to the effect of the Sun's
internal inhomogeneous magnetic field that necessarily rotates with
the ambient medium. To check this assumption, we have determined the
frequency for which the neutrino and magnetic data show the
strongest correlation by means of the joint spectrum statistic
\cite{ref:9}. This frequency is found to be 39.56 y$^{-1}$.
Figure~\ref{fig:9} shows the cumulative Rayleigh powers derived from
the neutrino and magnetic data for this frequency. We see that there
is a remarkable correspondence, both powers growing from 1996 until
1998,  then decreasing until about 2000.6 (solar maximum), after
which both increase sharply, indicating again the influence of the
solar cycle variations of magnetic structures in the convection
zone.

We now consider peak A at 9.43 y$^{-1}$ with power 11.51, and peak B
at 43.72 y$^{-1}$ with power 9.83. We presented evidence above for
r-mode oscillations \cite{ref:31} in the Homestake and GALLEX-GNO
data that appear to be the origin of the well-known Rieger
oscillation \cite{ref:21} and of similar ``Rieger-type" oscillations
\cite{ref:32} that have been discovered in recent years.  We have
therefore examined the possibility that peaks A and B may be related
to such oscillations interpreted as r modes.

R-modes are retrograde waves that, in a uniform and uniformly rotating model Sun,
have frequencies
\begin{equation}
\nu(\ell,m,{\rm syn})=m(\nu_R-1)-\frac{2m\nu_R}{\ell(\ell+1)}
\label{eq:2}
\end{equation}
as seen from Earth, where $\ell$ and $m$ are two of the usual spherical-harmonic indices, and $\nu_R$
is the sidereal rotation frequency (in cycles per year). If r-mode oscillations are to influence the solar
neutrino flux, they must interfere with magnetic regions that co-rotate with the ambient medium. The
interference frequencies will be given by
\begin{equation}
\nu=\left|m(\nu_R-1)-\frac{2m\nu_R}{\ell(\ell+1)}\pm m'(\nu_R-1)\right|
\label{eq:3}
\end{equation}
where $m'$, the azimuthal index for the magnetic structure, may be different from that of the r-mode.
For $m'=m$, equation~(\ref{eq:3}) yields the frequency
\begin{equation}
\nu(\ell,m,{\rm rot})=\frac{2m\nu_R}{\ell(\ell+1)},
\label{eq:4}
\end{equation}
for the minus sign, and
\begin{equation}
\nu(\ell,m,{\rm alias})=2m(\nu_R-1)-\frac{2m\nu_R}{\ell(\ell+1)}
\label{eq:5}
\end{equation}
for the plus sign. The frequencies given by (\ref{eq:4}) are the
well-known Rieger-type frequencies, which we showed in Sect.~2 were
found in the Cl and Ga data. We may regard the frequencies given by
(\ref{eq:5}) as aliases of these frequencies.

\begin{figure}
\epsfig{file=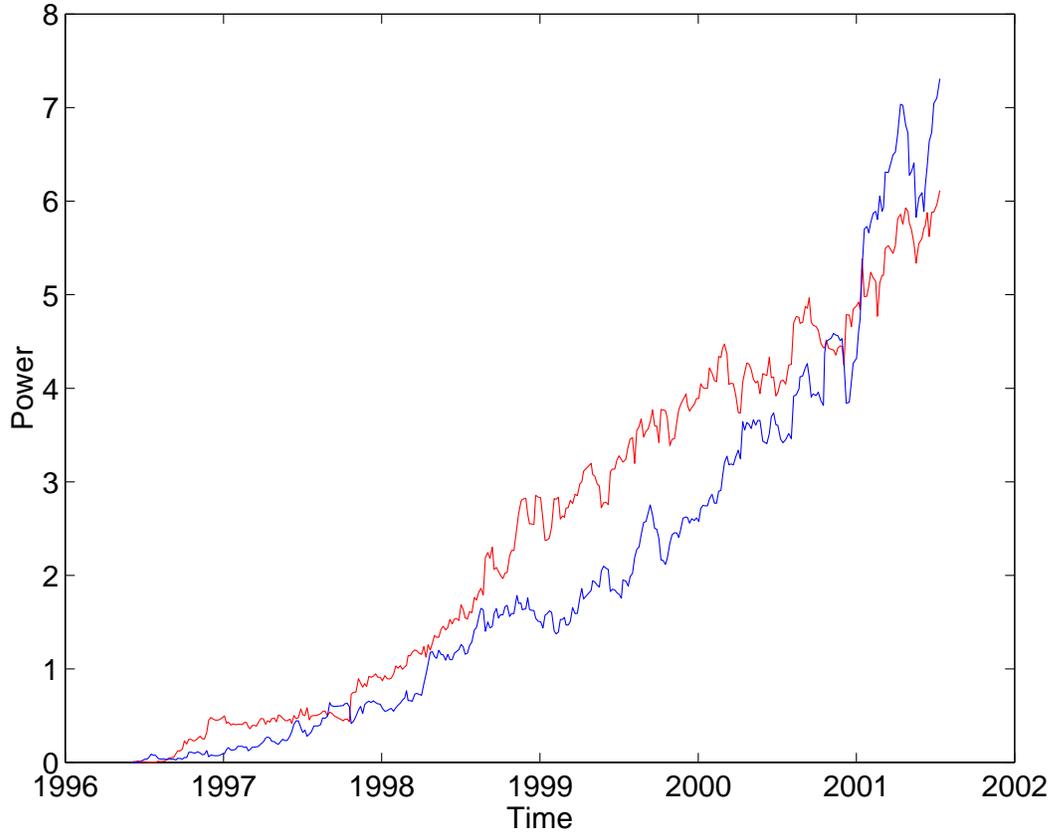,width=5.5in} \caption{Cumulative
contribution to the final Rayleigh powers of oscillations at 9.42
y$^{-1}$ (red) and 43.72 y$^{-1}$ (blue). \label{fig:10}}
\end{figure}

For $\ell = 2$ and $m = 2$ and for the range of values of $\nu_R$
inferred from the magnetic-field data, equation~(\ref{eq:4}) leads
us to expect oscillations in the band $9.47\pm0.09$, and
equation~(\ref{eq:5}) leads us to expect oscillations in the band
$43.33\pm0.47$. We see that the peaks A and B fall within these
bands. On applying the shuffle test, we find only 2 cases out of
10,000 in which a peak with power larger than 11.51 occurs in the
band $9.47\pm0.09$, and only 3 cases out of 1,000 that yield a power
larger than 9.83 in the band $43.33\pm0.47$. Figure~\ref{fig:10}
shows the cumulative Rayleigh powers at these two frequencies as a
function of time.  These two power components have a very similar
trend in their time evolution, indicating that they both originate
in the same internal oscillation, an r-mode with $\ell = 2$, $m =
2$.  Note the contrast between Figs.~\ref{fig:9} and \ref{fig:10},
showing the different origins of the A and B peaks from that of C.

\section{Discussion of the Frequency Peaks}
In the previous Section we have shown that frequency peaks A (9.43
y$^{-1}$) and B (43.72 y$^{-1}$) are not only related to each other,
but also that they can be quantitatively explained as r-mode
oscillations when one determines the rotation rate of the solar
magnetic field from observations. Frequency peak C (39.28 y$^{-1}$)
was shown to be the same as the third harmonic of the magnetic field
rotation rate, the only harmonic in the search band which was
sufficiently prominent, since the fundamental and second harmonic
were weak. Peak C illustrates the difficulty in predicting which
frequency peaks will be important, since at the 2000.6 solar maximum
the field broke up into structures which made third and seventh
harmonics of the rotation rate especially strong, as shown in
Fig.~\ref{fig:8}.

These same three peaks are seen in the analysis of the
Super-Kamiokande group \cite{ref:14a}, but they claim them not to be
statistically significant. In another paper \cite{ref:38} we have
addressed this disagreement in detail, but some of the results will
be summarized here. All comparisons are made standardizing on the
5-day data, a 0 to 50 y$^{-1}$ search band, and a Monte Carlo method
to provide the statistical significance of the strongest peak. With
the analysis method used by Super-Kamiokande, the Lomb-Scargle
procedure with the mean live times, the leading peak has a power of
7.29 at frequency 43.73 y$^{-1}$. Since 32\% of the simulations have
power equal to or greater than 7.29, the result is clearly not
statistically significant, in agreement with the conclusion of Yoo
et al.\ \cite{ref:14a}. If instead one uses a modified Lomb-Scargle
analysis \cite{ref:25}, employing again the mean live time but now
taking into account the average statistical error of each
data-point, the most significant peak becomes that at 9.43 y$^{-1}$
with a power of 9.56, which is significant at the 95\% confidence
level. This analysis still assumes that each data-point occurs at a
unique time. By using the likelihood analysis one can take into
account for each run the start time, end time, and mean live time,
as well as the statistical errors. This increases the power of the
9.43 y$^{-1}$ peak to 11.67, which has a significance level of
99.3\%. Clearly, by using this analysis method, we reach a different
qualitative conclusion than do Yoo et al.\ \cite{ref:14a}.

As more experimental information is used, the significance of the
flux modulation increases, as should be the case for a real signal.
Yoo et al.\ \cite{ref:14a} found from a sequence of Lomb-Scargle
analyses using simulated data with a sinusoidally varying neutrino
flux that, for periods of 20 days or more, they could not identify a
signal if the depth of modulation is less than 10\%. Since we find
the depth of modulation at 9.43 y$^{-1}$ (period 38.73 days) is only
7\%, we agree that they could not identify it reliably, whereas our
more sensitive analysis method does.

\section{Significance for Neutrino Physics}
The evidence given above for neutrino flux variability in the Cl,
Ga, and Super-Kamiokande data indicates that our current
understanding of either the Sun or the neutrino model is inadequate.
Changes in solar physics such as non-steady or non-spherically
symmetric nuclear burning would be difficult to reconcile with the
neutrino flux frequencies observed, or their apparent location in
some cases in the solar convection zone. Association of these
frequencies with the solar magnetic field rotation rates points very
strongly to their cause involving the
Resonant-Spin-Flavor-Precession (RSFP) process. This can occur,
however, only subdominantly to the MSW effect. The use of RSFP would
be simplest with the three known light active neutrinos, the mass
differences of which force the process to take place in the solar
core (from 0.05 $R_\odot$ at neutrino energy 3.35 MeV to 0.20
$R_\odot$ at 13 MeV \cite{ref:16new}), with the MSW effect for each
energy at a slightly larger radius. There are several disagreements
between this solution and the results of our analyses. First, flux
modulations show in all cases but one frequencies corresponding to
solar rotation rates in the convection zone above the tachocline
($\sim0.7R_\odot$). See Fig.~\ref{fig:1}, for instance. Second, the
depth of modulation in this model is too small. In the
Super-Kamiokande energy region the predicted \cite{ref:16new}
modulation is $\sim2$--3\%, but we find about 7\%. Third, the depth
of modulation is predicted \cite{ref:16new} to increase with
neutrino energy, being 0.77\% at 2.50 MeV and 1.03\% at 3.35 MeV,
whereas we find large effects at low energy; see Fig.~\ref{fig:4},
for example.

The next simplest model, which we introduced reluctantly some time
ago, now seems forced upon us. Already the large transition magnetic
moment required for RSFP necessitates new physics, so one hopes that
the need for a sterile neutrino arises from the same new physics.
The sterile neutrino is required for a fit to the data, for example
to provide $\Delta m^2\sim10^{-8}$ eV$^2$ needed to have RSFP in the
convection zone. Such a mass difference was essential even to fit
the time-averaged solar data when RSFP alone was used
\cite{ref:13,ref:14}. For the same reason that RSFP gave such a good
fit to the data (i.e., the energy dependence of the neutrino
survival probability), invoking it subdominantly in this sterile
neutrino model solves \cite{ref:10} two current problems with the
fits that occur when MSW alone is used. In addition to these
positive qualities, the usual negative ones are avoided. The lack of
solar antineutrinos places no limitation on RSFP because of the
sterile final state. Constraints from unitarity using atmospheric
and solar neutrino data are avoided, since the sterile neutrino does
not mix with active ones. For the same reason nucleosynthesis
considerations provide no limits. There is also sufficient
uncertainty in the fraction of solar neutrinos oscillating into
active neutrinos in the $^8$B energy region to allow for the needed
quantity of sterile final states \cite{ref:39}. Finally, it has been
shown \cite{ref:10} that this model can accommodate the depths of
modulation required. For a reasonable magnetic field shape (profile
2 in \cite{ref:10}), the maximum modulation depths are found to be
about 15\% in the Super-Kamiokande energy region, 17\% for Cl, and
40\% for Ga. Thus in every respect investigated, the model of a
sterile neutrino coupled to the electron neutrino via only a
transition magnetic moment satisfies the empirical information.

\section{Predictions and Conclusions}

The analysis of Super-Kamiokande data, when combined with the
results of analyses of Cl and Ga data, yields a variety of evidence
for variability of the solar neutrino flux.  It should be stressed
that this evidence can in principle all be understood in terms of
known fluid-dynamic and solar-dynamic processes in the standard
solar model. In particular, the frequency peaks in the
Super-Kamiokande data are all explained.  These same peaks are seen
in the analysis by the Super-Kamiokande collaboration, but their
less sensitive method of analysis makes these peaks appear
insignificant, whereas our method, that uses more of the
experimental information, can reveal these small modulations
\cite{ref:38}.

Although it is normal---perhaps inevitable---in exploratory research
that one examines the results of an analysis and then seeks to
interpret the results in known terms, this is a process that one
would like to avoid. In mature research, one would begin with a
prediction, and then look to see if that prediction is borne out in
a new study.

This procedure will be difficult to implement within the context of
solar neutrino modulation, for several reasons: we have inadequate
information concerning the flows and magnetic fields deep in the
solar interior, we must hypothesize about the neutrino physics
involved, and we have inadequate understanding of the way in which
dynamical processes near the tachocline lead to observable
consequences at the photosphere and above. Although helioseismology
has greatly improved our knowledge of the solar interior in recent
years, solar physics is not yet at the stage that we can safely
infer, from observational data, the flows and magnetic structures
which occur deep in the solar interior.

However, one may at this stage consider two sources of periodic
modulation that are well established, rotational effects and
Rieger-type oscillations \cite{ref:21,ref:32}, and attempt to
identify search bands associated with these dynamical processes. We
may expect to find modulation corresponding to the synodic rotation
frequency in the solar equatorial plane ($13.20\pm0.14$ yr$^{-1}$)
or a low harmonic of this frequency, such as the second
($26.40\pm0.28$ yr$^{-1}$) or third ($29.60\pm0.42$ yr$^{-1}$). The
power resulting from modulation at a given amplitude decreases as
$(2\pi\nu D)^{-2}$ where $D$ is the average length of each time bin
\cite{ref:26}, so that high harmonics will be difficult to detect.
Aliases of the higher harmonics would be correspondingly suppressed.
Concerning the Rieger-type oscillations, we propose that they may be
attributed r-mode oscillations, specifically the $\ell=3$, $m=1$
mode ($2.20\pm0.02$ yr$^{-1}$), the $\ell=3$, $m=3$ mode
($6.50\pm0.07$ yr$^{-1}$), and either the $\ell=3$, $m=2$ mode or
the $\ell=2$, $m=1$ mode, which have the same frequency
($4.40\pm0.05$ yr$^{-1}$). Hence it would be reasonable to search
for these modes and also the $\ell=2$, $m=2$ mode ($9.47\pm0.09$
yr$^{-1}$).

We apply this simple approach to the Super-Kamiokande power
spectrum. Concerning the rotational frequency bands, we found a
signal with power $S=8.91$ in the third band. For the specified
search band, we found from the shuffle test that the probability of
obtaining this strong a peak by chance is 0.005. The probability of
obtaining this result by chance in three ``trials" (for the three
rotational bands being considered) is 0.015. Concerning the four
4-mode bands being considered, we find a peak in one of them
($\ell=2$, $m=2$) with power $S=11.51$. We find from the shuffle
test that the probability of obtaining this strong peak by chance in
the relevant search band is 0.003. The probability of obtaining this
result by chance in four ``trials" for the four 4-mode bands being
considered is 0.012.

We now need to estimate the probability of obtaining both results in
the same dataset. From the relation $P=e^{-S}$, we find that
p-values of 0.015 and 0.012 convert to power values of 4.20 and
4.42, respectively. The corresponding combined power statistic
\cite{ref:9} \begin{equation}G=Z-\ln(1+Z),\end{equation} where $Z$
is the sum of the powers (i.e., 8.62), has the value 6.36. Hence
there is a probability of 0.0017 of obtaining these two peaks in
these seven search bands by chance.

This small probability does not utilize all the information in
Fig.~\ref{fig:8}, which shows why only the one harmonic found in the
neutrino data could be observed, provided the photospheric magnetic
field represents that in the solar interior. There is no similar
guidance in the r-mode case, since too little is known about the
complex solar interior ``weather" to predict which of the peaks
should be observable. For a different neutrino energy and solar
cycle, the other three r-modes were observed in the Cl and Ga
experiments, as described toward the end of Sec.~2. Concerning the
``alias" r-mode frequencies given by Eq.~\ref{eq:5}, we should
comment that, to the best of our knowledge, these have not
previously been identified in solar data, so it will be interesting
to see if they show up in subsequent neutrino data.

More precise specification of expected frequencies does not appear
possible at this time. That is why the power spectrum analysis of
the SNO data, expected shortly, is so important. The SNO and
Super-Kamiokande experiments had some overlap in time of operation
and observed a similar neutrino energy range. If one-day bins are
used, the alias peak at 62.56 y$^{-1}$ should not be seen, but the
r-mode frequencies of Fig.~\ref{fig:10} at 9.43 y$^{-1}$ and 43.72
y$^{-1}$ should be. In data taken in the solar cycle after 2000.6,
the 39.3 y$^{-1}$ frequency of Fig.~\ref{fig:9} should be present,
and with shorter sampling time, some of the higher harmonics of the
magnetic field rotation may also be observed.

Other types of tests exist of the RSFP framework, which appears to
explain so well the current information on solar neutrino flux
variations. First, experiments are being instrumented to measure
electron neutrino magnetic moments as low as $\sim10^{-12}\mu_B$,
which is in the range of interest. Second, future measurements of
the $^7$Be neutrino flux are likely to give a much lower value of
the time-averaged flux than predicted by the LMA solution.

The potential explanation for the cause of solar neutrino flux
variability makes proof of the effect all the more important. In
view of the results of the KamLAND experiment, a subdominant RSFP
process appears to be required, and this necessitates coupling the
electron neutrino to a sterile neutrino via only a large transition
magnetic moment. Such new physics, which appears compatible with all
known limitations and observations, should have many significant
consequences.

\section{Acknowledgments}
We are indebted to G.M.~Fuller, J.D.~Scargle, G.~Walther,
M.S.~Wheatland, and S.J.~Yellin for assistance and helpful
discussions. D.O.C.\ has been supported in part by a grant from the
Department of Energy No.~DE-FG03-91ER40618, and P.A.S.\ is supported
by grant No.\ AST-0097128 from the National Science Foundation.

\gdef\journal#1, #2, #3, #4#5#6#7{      
    {\sl #1~}{\bf #2}, #3 (#4#5#6#7)}   
\def\apj{\journal Astrophys.\ J., }
\def\app{\journal Astropart.\ Phys., }
\def\ass{\journal Astrophys.\ Space Sci., }
\def\baas{\journal Bull.\ Am.\ Astron.\ Soc., }
\def\nature{\journal Nature, }
\def\nc{\journal Nuovo Cimento, }
\def\np{\journal Nucl.\ Phys., }
\def\npps{\journal Nucl.\ Phys.\ (Proc.\ Suppl.), }
\def\pl{\journal Phys.\ Lett., }
\def\pr{\journal Phys.\ Rev., }
\def\prc{\journal Phys.\ Rev.\ C, }
\def\prd{\journal Phys.\ Rev.\ D, }
\def\prl{\journal Phys.\ Rev.\ Lett., }
\def\sjnp{\journal Sov.\ J.\ Nucl.\ Phys., }
\def\solarphys{\journal Solar Phys., }
\def\jetp{\journal J.\ Exp.\ Theor.\ Phys., }

\appendix{


\section{Appendix: Brief Explanation of the Likelihood Calculation}
The log-likelihood that the data $x_r$ may be fit by a given
functional form $X_r$ is given \cite{ref:26} by
$$L=-\frac{1}{2}\sum^R_{r=1}\frac{(x_r-X_r)^2}{\sigma^2_r},\eqno(1)$$
apart from a constant term, where $r=1,\ldots,R$ enumerates the
data bins. We adopt $$x_r=\frac{g_r}{{\rm mean}(g_r)}-1,\eqno(2)$$
where $g_r$ are the flux measurements, and scale the error
estimates accordingly.  We adopt the functional form
$$X_r=\frac{1}{D_r}\int^{t_{er}}_{t_{sr}}dt(Ae^{i2\pi\nu t}+
A^*e^{-i2\pi\nu t}),\eqno(3)$$ where $D_r$ is the duration of each
bin. Then, for each frequency, we adjust the amplitude A to
maximize the likelihood \cite{ref:3}.

What is significant is not the actual log-likelihood as computed
from (1), but the increase in the log-likelihood over the value
that corresponds to a constant flux, which is given (apart from
the same constant term) by
$$L_0=-\frac{1}{2}\sum^R_{r=1}\frac{x^2_r}{\sigma^2_r}.\eqno(4)$$
Hence the relative log-likelihood, which is equivalent to the
power spectrum, is given by
$$S=L-L_0,\eqno(5)$$
i.e., by
$$S=\frac{1}{2}\sum^R_{r=1}\frac{x^2_r}{\sigma^2_r}-\frac{1}{2}\sum^R_{r=1}
\frac{(x_r-X_r)^2}{\sigma^2_r}.\eqno(6)$$ Application of the
formalism above to one frequency in the Super-Kamiokande data
shows the magnitude of the flux modulation.  For $\nu=9.43$, and
for zero time chosen arbitrarily to be the date 1970.00, we find
that $A=0.0192-0.0291i$. This corresponds to a depth of modulation
of the neutrino flux of 7\%. }

\end{document}